# Tuning the critical temperature of cuprate superconductor films using self-assembled organic layers

I. Carmeli[*][1], A. Lewin[2], E. Flekser[2], I. Diamant[2], Q. Zhang[3], J. Shen[3], M. Gozin[1], S. Richter[*][1], and Y. Dagan[*][2]

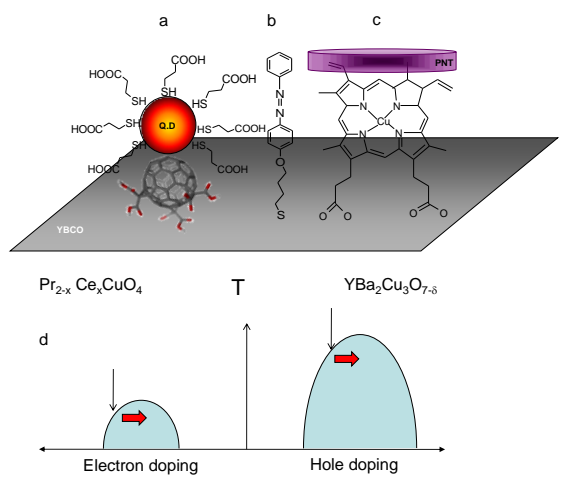

**Scheme 1.** Schematic illustration of the various molecular films deposited on the HTCS. a, CF monolayer coated with CdS. b, AZ monolayer. c, Heterostructure SAM composed of porphyrin nanotubes (PNT) condensed on porphyrin monolayer. For a CF monolayer, upon addition of CdS, an increase in $T_c$ is observed. Reversible $T_c$ modulation has been achieved by an AZ monolayer and the porphyrin-based heterostructure. d. Phase diagram of the two superconductors used. The horizontal axis is the carrier concentration (n) or doping. The origin represents the undoped insulating compound. Moving to the left (or right) represents increasing electron (or hole) doping. Superconductivity is observed in the dome-shaped regions of this doping-temperature phase diagram. The shift on the phase diagram resulting from the SAM for both electron- (overdoped) and hole-doped (underdoped) cuprates is indicated by the red arrow. The SAM extracts electrons from both types of surfaces and this results in an increase in $T_c$.

Many of the electronic properties of high-temperature cuprate superconductors (HTSC) are strongly dependent on the number of charge carriers put into the $CuO_2$ planes (doping). Superconductivity appears over a dome-shaped region of the doping-temperature phase diagram. The highest critical temperature ($T_c$) is obtained for the so-called "optimum doping". The doping mechanism is usually chemical; it can be done by cationic substitution. This is the case, for example, in $La_{2-x}Sr_xCuO_4$ where $La^{3+}$ is replaced by $Sr^{2+}$ thus adding a hole to the $CuO_2$ planes. A similar effect is achieved by adding oxygen as in the case of $YBa_2Cu_3O_{6+\delta}$ where $\delta$ represents the excess oxygen in the sample. In this paper we report on a different mechanism, one that enables the addition or removal of carriers from the surface of the HTSC. This method utilizes a self-assembled monolayer (SAM) of polar molecules adsorbed on the cuprate surface. In the case of optically active molecules, the polarity of the SAM can be modulated by shining light on the coated surface. This results in a light-induced modulation of the superconducting phase transition of the sample. The ability to control the superconducting transition temperature with the use of SAMs makes these surfaces practical for various devices such as switches and detectors based on high-$T_c$ superconductors.

In cuprate superconductors, $T_c$ has an approximately quadratic dependence on carrier concentration.[1] The carrier concentration is an order of magnitude lower than that of metals and the screening is weaker. This raises the possibility of modifying the carrier concentration by applying an electric field to the surface. One approach was to use field-effect devices.[2, 3] Here we report on an alternative approach to doping the surface of the HTSC, namely with the use of a molecular self-assembly method.

A self-assembled monolayer (SAM) is an organized layer of molecules in which one end of the molecule, the binding group, is designed to interact favourably with the solid surface of interest, forming a well-organized monolayer on it.[4] The SAM is terminated with a functional group - in our case these are surface dipole-forming molecules or optically active moieties. The expected change in the work function due to a dipole moment of 1 Debye per molecule in a "typical" SAM with molecular density of $5\times10^{14}$ $cm^{-2}$ is ~0.5eV. The resulting electric field is compensated by charge transfer between the substrate and the SAM. The amount of charge transfer per molecule is $q=\mu/r$, where $r$ is the length of the molecule. When adsorbed on a cuprate surface, this charge transfer can be designed to induce holes and thus change the surface carrier density and the resulting critical temperature. In the case of photo-active molecules this change in $T_c$ can be induced by shining light on the SAM *via* electron-hole-pair excitation. In earlier work, several groups studied SAM adsorption on YBCO but found no change in the critical temperature after adsorption of various SAMs on the superconducting surface.[5]

In this study, several types of molecular monolayers and heterostructure were designed. These compounds exhibit strong charge-transfer characteristics and photo response when adsorbed on cuprate surfaces. In order to investigate the effect of SAMs on cuprates, three types of representative films were chosen, each targeted to a specific response upon adsorption or light excitation.

The first monolayer consists of carboxyfullerene (CF) coupled to CdS Quantum dot (CF/QD, see scheme(a)). In general, CF can be reduced by one electron per molecule. This reduction can occur

[*]  I. Carmeli, M. Gozin, S. Richter
Raymond and Beverly Sackler School of Chemistry
E-mail: (itai@post.tau.ac.il, srichter@post.tau.ac.il)
A. Lewin, E. Flekser, I. Diamant, Y. Dagan
Raymond and Beverly Sackler School of Physics and Astronomy
Tel-Aviv University
Tel Aviv, 69978, Israel
Fax: (+972-3-6408286)
E-mail: yodagan@post.tau.ac.il

Q. Zhang, J. Shen
Shanghai Institute of Materia Medica (SIMM)
Chinese Academy of Sciences
501 Haike Road, Shanghai 201203, China

[**] Support from the Israel Science Foundation under grants1421/08 and the Israeli Ministry of Science and Technology is acknowledged

upon application of a gate voltage.[6] In our case, the gate voltage is replaced by a CdS over-layer which reduces the charge in the CF monolayer, thus forming an interfacial dipole between the two layers. This structure is used to study the effect of an interfacial dipole on $T_c$.

An Azobenzene derivative (1-Butanethiol,4[4-(phenylazo)phenoxy] (AZ), Scheme(b)), an optically active molecule, was used to modulate $T_c$ by light. This molecule exhibits a reversible light-induced conformational change which results in two distinct dipole states (-0.15D and +1.53D for trans- and cis-conformations, respectively). The AZ monolayer has been shown to modulate the magnetism of thin gold surfaces[7] and change the transport and superconducting properties of thin niobium films.[8] Here we demonstrate the ability of an AZ monolayer to modulate the transition temperature of high-$T_c$ superconductors.

Next, we studied porphyrin-based SAMs coupled to porphyrin nanotubes (Scheme(c)). Light-induced charge transfer between porphyrins and carbon nanostructures,[9] or metals[10] has been demonstrated. In the latter case, a photopotential of several tenths of a millivolt has been measured. In our study we exploited the special properties of porphyrin nanotubes, which, upon illumination, are expected to induce holes in the superconductor.

Scheme (d) shows the phase diagram of the cuprate superconductors used: the hole-doped $YBa_2Cu_3O_{6+\delta}$ (YBCO) and the electron-doped $Pr_{2-x}Ce_xCuO_4$. While the parent compounds are antiferromagnetic insulators, upon addition of charge carriers to the $CuO_2$ planes, the material becomes superconducting with a critical temperature that depends on the carrier concentration, reaching its maximum at the so-called optimum doping; increasing the carrier concentration further results in a decrease in $T_c$.

All our SAMs were tested on YBCO, which has a $T_c$ over a temperature range accessible with liquid nitrogen. For a further demonstration of the physical effect that takes place at the surface, the CF/QD system was adsorbed on PCCO which is an electron-doped cuprate superconductor[11] and on YBCO. The black arrows in the scheme depict the carrier concentrations used in this experiment. YBCO underdoped and optimally doped samples were used, while for PCCO we used x=0.17 which is in the overdoped regime. For both underdoped YBCO and overdoped PCCO, the removal of electrons from the surface results in an increase in $T_c$ (shift indicated by the rightward-pointing red arrows in scheme (d)).

CF is known to form a dense monolayer when adsorbed on oxide surfaces.[12] We found that monolayers of similar quality are formed when CF is adsorbed on cuprates, as indicated by the large contact angle measured (see experimental section). It is important to note that, upon addition of the QD layer, the dipolar field reverses its orientation, as inferred from the corresponding changes in $T_c$. This reversal in the dipolar field is also evident from Kelvin Force Microscopy (KFM) measurements (see supplementary).

Figure 1 shows the dependence of resistance on temperature for PCCO (left-hand panel) and YBCO (right-hand panel) and the corresponding reference samples. $T_c$ decreased when a dense layer of CF was adsorbed on both surfaces. A remarkable increase in $T_c$ is observed for both the electron- and hole-doped cuprates when an additional layer of cadmium sulfide is added. The observed enhancement of $T_c$ corresponds to a charge transfer of ~0.1 carriers per unit cell. This is consistent with the change in surface potential found from KFM measurements (See supplementary).

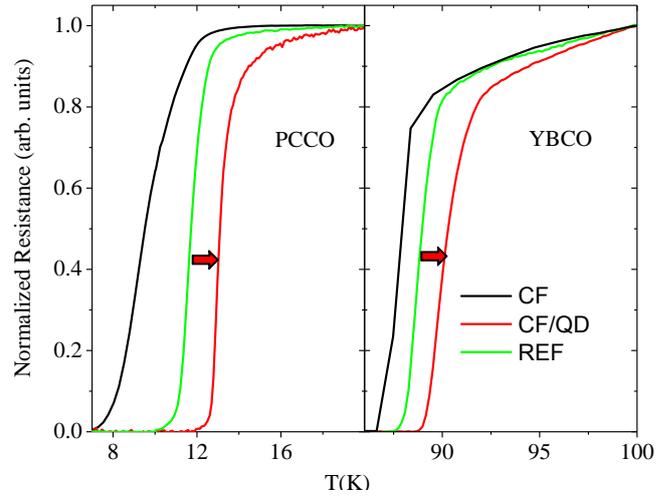

**Figure 1.** Normalized resistance measurement of CF coupled to CdS quantum dot on $Pr_{1.83}Ce_{0.17}CuO_4$ (left) and on $Y_1Ba_2Cu_3O_{7-\delta}$ (right) films. Bare film resistivities are about 80μΩ-cm at 100K for both samples. Adsorption of the CF SAM results in a decrease in $T_c$ (black) compared to the uncoated reference sample (green) and CF/QD coated one (red line) where a substantial $T_c$ enhancement is observed for both types of cuprates.

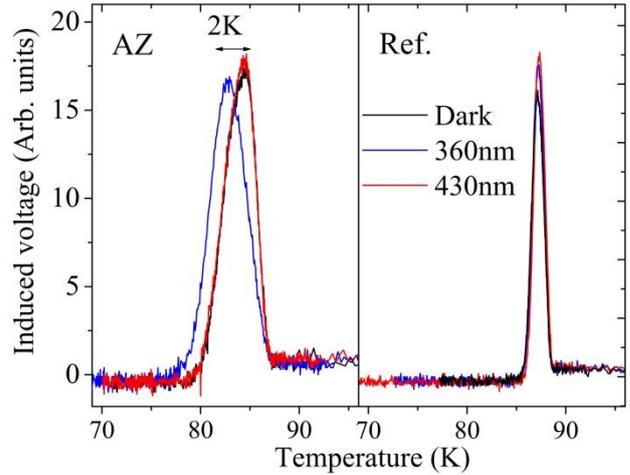

**Figure 2.** Reversible light-induced $T_c$ switching of YBCO-coated AZ monolayer (left panel) compared to the null effect found in the SAM-free reference sample (right panel) measured by the imaginary component of the induced voltage in the AC susceptibility. Excitation by UV light (360nm) causes a 2K decrease in $T_c$ while illumination with 430nm (red curve) increases $T_c$ to its original value (black curve). $T_c$ is defined by the peak in the signal. A light-induced $T_c$ modulation of 2K is observed.

Fig.2 (left) shows the light-dependence of the imaginary part of the AC susceptibility for a YBCO-coated AZ monolayer. In this measurement, the peak of the signal corresponds to $T_c$. A clear reversible light-induced switching of $T_c$ is observed compared to the null effect found in the SAM-free reference sample (see Fig.2 right panel). We attribute this effect to the conformational changes of the azobenzene ring accompanied by reversal of the molecule dipole induced by illumination (see supplementary and references therein). The long relaxation time of each molecular state may pave the way to zero-power switching devices and light-induced dissipationless memories.

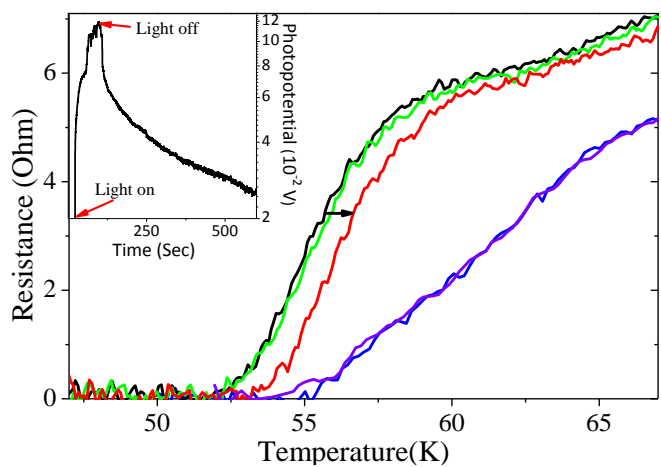

***Figure 3.*** $T_c$ modulation of YBCO sample with the use of a porphyrin-based heterostructure stimulated by a single light source. The $T_c$ of the sample increased by 1K when illuminated by 520nm diode light (red curve). The $T_c$ relaxes (green curve) back almost to its original value (black curve) once the light is turned off. No effect is detected for the reference sample (blue and violet curves for light on and off respectively). **Inset,** a photopotential of 120mV is generated by light excitation of the porphyrin heterostructure. This photopotential relaxes to zero when the light is turned off. Both the surface potential and its dynamics are in qualitative agreement with the $T_c$ modulation.

In Figure 3 we show $T_c$ modulation with the use of a porphyrin-based heterostructure stimulated by a single light source. As expected, illuminating the sample results in an increase in $T_c$. $T_c$ decreases to its original value when the light is switched off, indicating that the charge has been transferred back to the (hole-doped) cuprate layer. The change in carrier concentration inferred from the enhancement of $T_c$ and the charge dynamics is in agreement with time-resolved surface-potential measurements by Kelvin Force Microscopy (See inset Fig.3). This indicates that light induces electron transfer to the porphyrin heterostructure. These electrons relax back to the cuprate surface once the light is switched off.

We shall now discuss the possible mechanisms of $T_c$ modulation. The strong dipolar field created by the SAM is screened by charge transferred from the cuprate surface. In all cases described above, electrons are transferred from the superconductor, leaving a hole at the surface which increases (decreases) the extent of hole (electron) doping. This process results in $T_c$ enhancement for both underdoped YBCO and overdoped PCCO. A simple calculation of the charge transfer needed to compensate for the field, deduced from KFM surface-potential measurements assuming a SAM thickness of a few nanometers, gives the right order of magnitude, namely 0.01 holes per unit cell.

However, our results suggest that the charge transfer occurs over a distance greater than the coherence or the screening lengths (a few nanometers for both cases). Otherwise, it is difficult to explain the decrease in $T_c$ upon CF or AZ adsorption (see supporting information). Charge transfer on relatively large length-scales has been reported for electron-doped cuprates. [13] It is possible that this charge transfer, together with a proximity effect, results in $T_c$ enhancement in a significant volume fraction of the film.

Of the various SAMs tested here, it seems that the CF and the AZ SAMs are the most efficient. This is evident from the relatively large change in $T_c$ obtained for PCCO when using the former and a significant change in YBCO in the vicinity of the dome maximum. During the preparation of the manuscript we became aware that a $T_c$ enhancement of Nb thin films proximity-coupled to gold nanoparticles using organic linkers has been recently reported. [14]

The novel approach for modulating $T_c$, based on the design of functional SAMs presented here, may pave the way to new dissipationless memories and switches based on high-$T_c$ superconductors.

## Experimental Section

YBCO films 500Å thick were deposited by RF-sputtering at 770°C and 400mTorr oxygen pressure. For underdoped samples, the films were annealed at a low oxygen pressure of 0.1Torr for 3 hours. PCCO films were deposited by the pulsed-laser deposition technique at 780°C and 230mTorr $N_2O$. For all films used, one piece of the film was kept as a reference. It went through a process identical to that of the samples but without SAM adsorption.

Compounds A-C were synthesized by us and adsorbed as SAMs on the superconductor surfaces (see Supplementary Information section).

Measurements: Resistance was measured with the use of a modified Quantum Design PPMS platform that enabled us to shine light through an optical fibre during cool-down. AC susceptibility was measured by a lock-in technique with the use of two coils.

# Supplementary Information

# Tuning the critical temperature of cuprate superconductors using self - assembled organic layers

I. Carmeli [1], A. Lewin [2], E. Flekser [2], I Diamant [2], Q. Zhang [3], J. Shen [3], M. Gozin[1], S. Richter[1] and Y. Dagan[2]

[1] *Raymond and Beverly Sackler School of Chemistry, Tel-Aviv University, Tel Aviv, 69978, Israel*
[2] *Raymond and Beverly Sackler School of Physics and Astronomy, Tel-Aviv University, Tel Aviv, 69978, Israel*
[3] *Shanghai Institute of Materia Medica (SIMM), Chinese Academy of Sciences, 501 Haike Road, Shanghai 201203, China*

## 1. Photo-potential measurements using Kelvin Force Microscopy

Photo-potential of the samples was measured using Kelvin Force Microscopy (KFM). For this purpose we used Au coated TiN KFM tips which were calibrated to a reference graphite surface. The measurements were conducted using Solver PH47 (NTMDT Inc.) operating in tapping mode at the cantilever resonance frequency of around 130 kHz. The Solver PH47 was equipped with a custom-made 1300-nm wavelength feedback laser to prevent unwanted laser induced photo-voltage.

The surface potential of the bare YBCO, sample was found to be $450\pm20$ mV. The surface potential of sample A was found to be $580\pm20$ mV while the surface potential of the CF monolayer adsorbed on YBCO exhibited values of $310\pm20$ mV. The difference in work functions indicates that adsorption of CF causes a clear reduction of surface potential while topping the CF with CdS Q.D (sample A) induces charge transfer from the surface to the hybrid structure and increases the surface potential. No light effect on the surface photopotential was observed in these samples.

## 2. Dipole modulation induced by conformational changes in the Az molecule

The trans- and cis- conformations of the AZ molecules (sample B) are known to exhibit different dipoles. Thus, they can be used to modulate the work function of the substrate upon illumination.[1] The reported dipole values are -0.15 D and +1.53 D for trans and cis-conformations, respectively. Conformational changes are induced by illumination with UV (365nm) and white light.

## 3. Mechanism for charge transfer

Our results suggest that the charge transfer occurs on a length-scale larger than the coherence or the screening lengths (a few nanometers for both cases). Otherwise, it is difficult to explain the decrease in $T_c$ upon CF or AZ adsorption. If only the surface is affected by monolayer adsorption, which decreases its $T_c$, upon lowering the temperature below the bulk transition one would expect the resistive surface to be short-circuited by the superconducting bulk resulting in total zero resistance. Therefore, in this scenario such surface $T_c$ reduction will not be seen in a resistivity measurement. The fact that $T_c$ reduction is seen in our results suggests that the charge transfer occurs on a longer length-scale.

A deeper insight into the charge transfer profile can be attained from the AC susceptibility measurement presented in fig.3. The onset of the transition is not affected by the light stimulation suggesting that the deepest layers of the film are not affected by the SAM whereas the peak of the imaginary part moves upon light stimulation. The shift in the peak position corresponds to a $T_c$ change in a large volume fraction in the sample.

Oxygen diffusion may screen the surface dipole and result in a longer length scale for $T_c$ enhancement. Oxygen plays an important role as a dopant for both YBCO and PCCO. See main text for further discussion.

# 4. SAMs synthesis and characterization

**Compound A:**

*Synthesis of the Carboxy fullerenes (CF)* [2]

Diethyl bromomalonate was added to a solution of C60 in toluene, followed by the addition of 1,8-diazo-bicyclo[5,4,0]undec-7-ene. After stirring for 4 days, the solvent was removed in vacuo, and the residue was chromatographed on silica gel using toluene as eluent. The third band corresponded to semipure C3 as the major component. The fractions containing semipure C3 or D3 were rechromatographed on silica gel (230 – 400 mesh) using toluene as the mobile phase to give the purified C3 or D3 isomers. To a solution of either C3 or D3 (100 mg in 100 ml toluene, 0.1 mmol) was added NaH (80%, 60 mg, 2 mmol), and the mixture was refluxed for 1 hr. After the heating source was removed, MeOH (5 ml) was added immediately to quench the reaction. Red powder precipitated and was collected by centrifugation. The powder was washed with toluene and hexanes, then dissolved in water to which HCl (4 M) was added. A red amorphous precipitate was formed immediately, which was collected by centrifugation. The solid again was washed with HCl (4 M), and then with water. The solid was dissolved in MeOH, and the solvent was removed in vacuo to give the powdery pure red C3 isomer acid.

*Self-assembly of compound A*

First the CF were self-assembled on the superconducting surface from a solution of 1mM CF in Ethanol over-night. The samples were than washed in pure Ethanol and dried in $N_2$. Samples were characterized by contact angle, measuring contact angle of 100°, indicating that a well-organized layer was formed. The CF monolayer was than coated with Mercapto propionic acid modified (according to the procedure described in ref 3) CdS quantum dots (MK nano MKN-CdS-440) by drop casting 2μL of 100nM solution of top of the CF monolayer.

**Compound B:**

The synthesis route is shown in the scheme below:

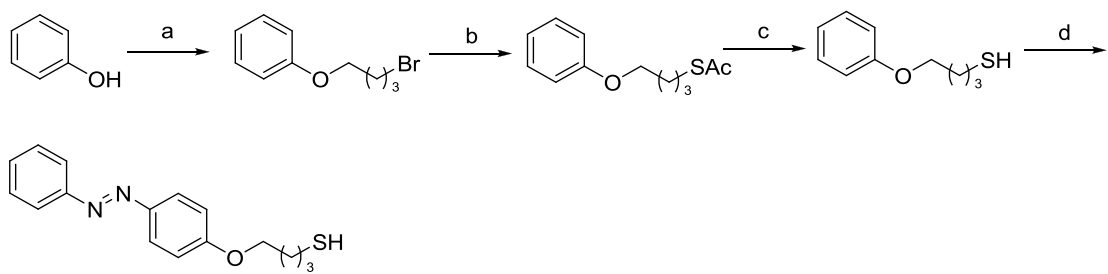

*Reagents and Conditions*: a) 1,4-dibromobutane, $Cs_2CO_3$, DMF, 75%; b) KSAc, DMF, 96%; c) NaOH, $H_2O$, 95%; d) aniline, $NaNO_2$, conc. HCl, $H_2O$; then 4-phenoxybutane-1-thiol, NaOH, $H_2O$, 88%.

Next compound B were adsorbed and dried from a solution of 1mM in Ethanol giving contact angle of ~90°.

**Compound C:**

Coupling SAM of porphyrins to porphyrin nanotubes was achieved by immersing the cuprate film in a 0.5mM solution of the porphyrins (Frontier Scientific Cu(II) protoporphyrinIX P40769) dissolved in 200μL of pyridine, and 1ml THF. Than 1ml Hexane and Cyclohexane were added to the solution resulted in condensation of the porphryns into nanorods typically 70nm in diameter (as indicated by AFM imaging).

# 5. Properties of $Y_1Ba_2Cu_3O_7$ and $Pr_{1.83}Ce_{0.17}CuO_4$

The films were studied by x-ray diffraction and scanning electron microscopy. The results are described elsewhere: YBCO and PCCO properties are summarized in Refs. [4,5] respectively.

The resistance as a function of temperature is shown below for YBCO and PCCO samples with and without SAM adsorption. The film thickness is about 40nm and resistance is measured roughly in a square geometry. The YBCO data are raw data of Fig. 1. The PCCO data are of an additional sample with a slightly higher $T_c$. The $T_c$ enhancement effect for this sample is similar. It is noted that the small resistance decrease seen in both cuprates is expected for the YBCO sample since increase of carrier concentration results in decrease in resistivity. However it cannot be understood for PCCO where the resistance is expected to increase. We can attribute this effect to a small contribution of the SAM itself to the conductivity and possibly to a slight difference in contact configuration. It is however possible that doping by charge transfer has a different effect on the resistivity comparing to chemical substitution. Such a behavior has been recently reported for $T_c$ modulation using the field effect on YBCO samples.[6]

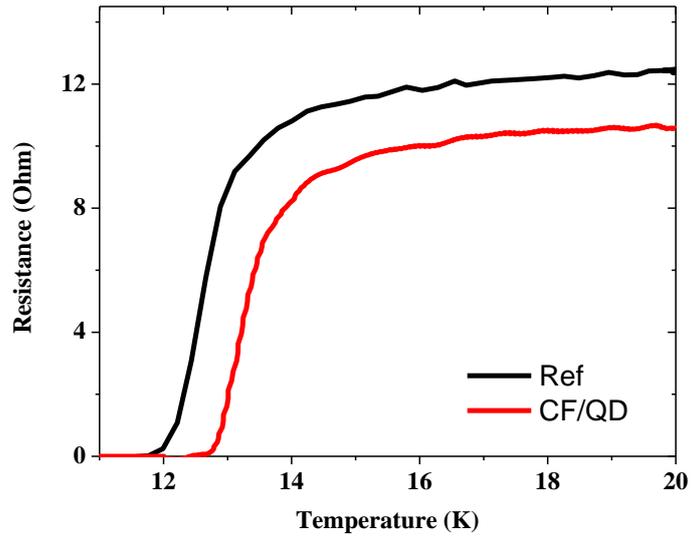

**PCCO sample with and without SAM**

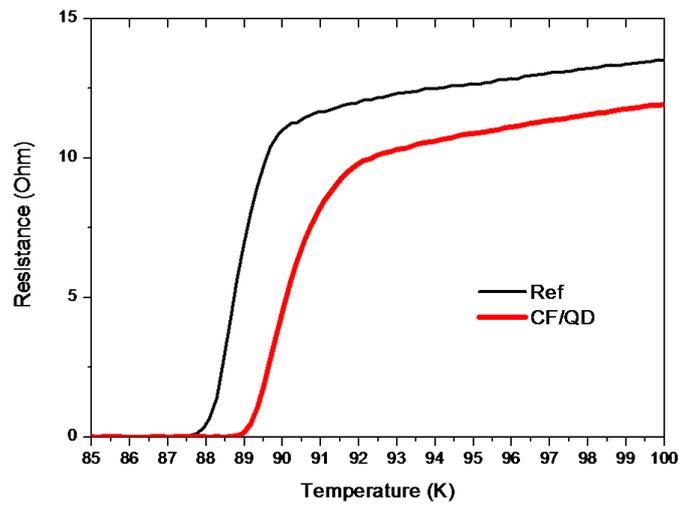

**YBCO film with and without SAM**

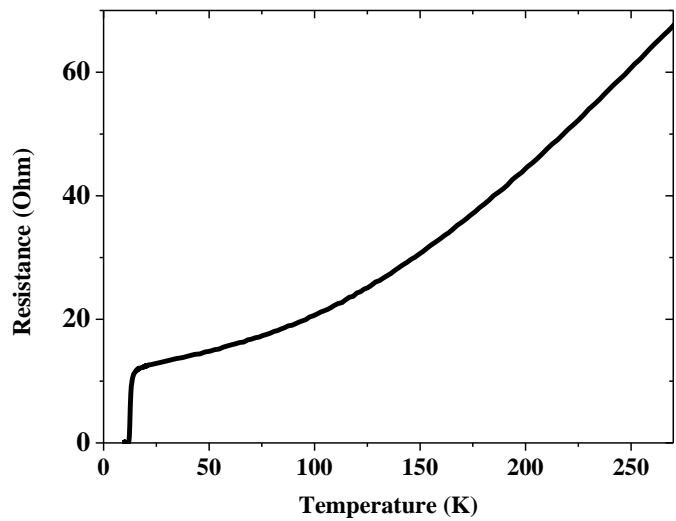

**PCCO Reference sample**